\documentclass[prb,twocolumn,floatfix]{revtex4}

\usepackage{graphicx}
\usepackage{dcolumn}
\usepackage{bm}

\newcommand{\be}{\begin{equation}}
\newcommand{\ee}{\end{equation}}
\newcommand{\ba}{\begin{eqnarray}}
\newcommand{\ea}{\end{eqnarray}}
\newcommand{\nn}{\nonumber \\}
\def\ket#1{\left\vert #1 \right\rangle}

\begin{document}
\title{Non-Markovian incoherent quantum dynamics of a two-state system}

\author{M.~H.~S.~Amin}
\author{Frederico Brito}
\affiliation{D-Wave Systems Inc., 100-4401 Still Creek Drive,
Burnaby, B.C., V5C 6G9, Canada}

\begin{abstract}
We present a detailed study of the non-Markovian two-state system
dynamics for the regime of incoherent quantum tunneling. Using
perturbation theory in the system tunneling amplitude $\Delta$, and
in the limit of strong system-bath coupling, we determine the short
time evolution of the reduced density matrix and thereby find a
general equation of motion for the non-Markovian evolution at longer
times. We relate the nonlocality in time due to the non-Markovian
effects with the environment characteristic response time. In
addition, we study the incoherent evolution of a system with a
double-well potential, where each well consists several quantized
energy levels. We determine the crossover temperature to a regime
where many energy levels in the wells participate in the tunneling
process, and observe that the required temperature can be much
smaller than the one associated with the system plasma frequency. We
also discuss experimental implications of our theoretical analysis.
\end{abstract}

\maketitle

\section{Introduction}
It is difficult to overemphasize the importance of the dissipative
dynamics of a two-state system (TSS). In general, standing as a
first hand approximation of a much rather complex level structure,
the model of a TSS coupled to a dissipative environment
\cite{leggett,weiss} has been successfully applied to several
physical systems. Indeed, the dissipative TSS dynamics is the
paradigm for the study of superconducting devices containing
Josephson junctions,  \cite{super} two-level atoms in optical
cavities, \cite{optics} electron transfer in biological and chemical
systems \cite{electrontransfer} and semiconductor quantum dots,
\cite{qdots} to name just a few.

Despite its simplicity, the description of the TSS dissipative
dynamics imposes great theoretical challenges, especially when
considering non-Markovian processes. This is the case for the
analysis of the environment low-frequency noise spectrum, since the
long-lived feature of its fluctuations does not allow for a
``memoryless'' bath (Markov) approximation.  In the context of a
weak TSS-bath coupling, theoretical efforts have been made to
quantify the low-frequency effect for both spin-boson
\cite{divincenzo} and $1/\rm{f}$ noise\cite{guido,schriefl} models.

Furthermore, it has been largely demonstrated that low-frequency
noise plays important role in the decoherence process of
superconducting devices containing Jospehson
junctions.\cite{astafiev,muck,yoshihara,Kakuyanagi,bialczak} Since
those devices are seen as promising candidates to the physical
implementation of a quantum bit, this subject has rapidly grown in
interest and several studies on describing the microscopic origin
and characterizing the low-frequency noise in such devices have
already been put forward.\cite{roger,faoro,trevor,sendelbach}

Understanding the evolution of a TSS also plays an important role in
understanding the performance of an adiabatic quantum computer
\cite{Farhi} in the presence of noise.
\cite{Ashhab,ALT08,AAN09,ATA09} This is because for many hard
problems
the bottleneck of the computation is passing through a point where
the gap between the ground state and first excited state is very
small. Near such an energy anticrossing, the Hamiltonian of the
system can be truncated into a two-state Hamiltonian \cite{ATA09}
and in the regime of strong coupling to the environment the
two-state results discussed in this paper may be directly applied.

Here, following a previous work,\cite{aa08} we put forward a
detailed study of the TSS dissipative dynamics in the presence of
low-frequency noise, for the regime of strong TSS-bath coupling. We
show that, for the regime of small tunneling amplitude $\Delta$,
dephasing takes place much earlier in the evolution, leading the
system to incoherent quantum dynamics. We employ such a property to
derive equations that describe the non-Markovian evolution of the
system's density matrix.

The paper is organized as follows. In section
\ref{systemhamiltonian}, we present the system Hamiltonian and a
formal solution for the time evolution operator. Assuming second
order perturbation theory in $\Delta$, we calculate in section
\ref{densitymatrix} the short-time dynamics of the system reduced
density matrix elements. Section \ref{equilibrium}, presents a
discussion regarding the non-Markovian behavior of the system when
the environment is in equilibrium. We determine conditions under
which the system reaches the detailed balance regime. Section
\ref{eqmotion} provides a systematic derivation of an equation of
motion for the system evolution, which in general is non-local in
time. We also discuss regimes in which the equations governing the
diagonal part of the density matrix become $t$-local. Considering a
double-well potential, section \ref{doublewell} puts forward the
analysis of intra- and interwell transitions and situations where a
classical mixture of states participate in the quantum tunneling
process. Finally, section \ref{conclusion} presents our concluding
remarks.

\section{System Hamiltonian}
\label{systemhamiltonian}

We start by considering an open two-state system with Hamiltonian
\be
 H = -\frac{1}{2}[\Delta(t) \sigma_x + \epsilon(t) \sigma_z]
 - \frac{1}{2} \sigma_z Q + H_B, \label{HS}
\ee
where $Q$ is an operator acting on the environment described by the
Hamiltonian $H_B$.

In order to determine the system evolution operator $U(t_2,t_1)$, we
proceed through two simple steps. First, we write the state vector
of the system Hamiltonian (\ref{HS}) as $|\psi(t)\rangle=e^{iH_B
t}|\varphi(t)\rangle$. ($\hbar=k_B=1$, through this paper.) Thus,
one finds that the state vector $|\varphi(t)\rangle$ evolves in time
according to $i\frac{\partial}{\partial
t}|\varphi(t)\rangle=[H_0(t)+V(t)]|\varphi(t)\rangle$, where
 \be
 H_0(t)= -\frac{1}{2}\epsilon(t) \sigma_z - \frac{1}{2} \sigma_z Q(t),
 ~~V(t)=-\frac{1}{2}\Delta(t)\sigma_x, \label{H0t}
 \ee
and $Q(t) = e^{iH_B t}Qe^{-iH_B t}$. The environment is assumed to
feature fluctuations following Gaussian distribution, therefore all
averages can be expressed in terms of the correlation function or
its Fourier transform, the spectral density:
 \be
  S(\omega)=\int_{-\infty}^\infty dt \ e^{i\omega t} \langle Q(t) Q(0)
 \rangle,
 \ee
hence we do not need to specify $H_B.$\cite{footnote}

The next step is to make use of the interaction picture, considering
$V(t)$ as the perturbation. The state vector in the interaction
picture is defined by $|\varphi_I(t)\rangle \equiv U_0^\dagger (t)
|\varphi(t)\rangle$, and any operator $\hat O$ is transformed by
$\hat O_I(t) = U_0^\dagger (t)\hat O U_0 (t)$, with
 \ba
 U_0(t) &=& {\cal T} e^{-i\int_0^{t} H_0(t') dt'} \nn
 &=& {\cal T} \exp \left\{i\frac{\sigma_z}{2}
 \int_0^t [\epsilon(t')+Q(t')]dt' \right\}, \label{U}
 \ea
where ${\cal T}$ denotes the time ordering operator.
%
%
Now, the state evolution is determined by the interaction potential
 \ba
 H_I (t)
 = -\frac{1}{2}\Delta(t) \tilde{\sigma}_x(t),
 \ea
where $\tilde{\sigma}_x(t) = U_0^\dagger(t)\sigma_xU_0(t)$. The time
evolution operator in the interaction representation reads
 \ba
 U_I(t_2,t_1) = {\cal T} e^{-i\int_{t_1}^{t_2} H_I(t) dt}.
 \ea
Finally, we can write a formal solution for the complete time evolution operator as
 \ba
 U(t_2,t_1) &=& {\cal T} e^{-i\int_{t_1}^{t_2} H(t) dt} \nn
 &=& e^{-iH_B t_2}U_0(t_2)U_I(t_2,t_1)U_0^\dagger(t_1)e^{iH_B t_1}. ~~\label{S0t}
 \ea

In this paper, we are interested in the strong coupling regime in which the r.m.s.
value of the noise
 \be
 W \equiv \sqrt{\langle Q^2\rangle} = \left(\int_{-\infty}^\infty
 {d\omega \over 2\pi}S(\omega) \right)^{1/2}, \label{W}
 \ee
is much larger than the tunneling amplitude: $W \gg \Delta$.
Physically, $W$ is basically the uncertainty in the energy bias
$\epsilon(t)$ and therefore represents the broadening of the energy
levels. In the above regime, consequently, the broadening of the
energy levels is larger than the minimum gap and therefore the gap
will not be well-defined. On the other hand, as we shall see, for
the case of low frequency noise, $W$ represents the dephasing rate
of the system. Thus, the above regime is a limit in which the qubit
loses quantum coherence before it can tunnel, i.e., the dynamics is
incoherent.

\section{Density matrix calculation}
\label{densitymatrix}

We would like to study the evolution of the reduced density matrix.
Let $\rho_{SB}(t)$ denote the total density matrix of the system
plus bath. We therefore have
 \ba
 && \rho_{SB}(t) = U(t,0)\rho_{SB}(0)U^\dagger(t,0) \\
 && = e^{-iH_B t}U_0(t)U_I(t,0)\rho_{SB}(0)
 U_I^\dagger(t,0)U_0^\dagger(t)e^{iH_B t}. \nonumber
 \ea
The system reduced density matrix is defined by $\rho(t)=\text{Tr}_B
[\rho_{SB}(t)]$, where Tr$_B[...]$ means averaging over all
environmental modes. We assume that the density matrix at $t=0$ is
separable, i.e., $\rho_{SB}(0) =\rho(0)\otimes\rho_B $, where
$\rho_B = e^{-H_B/T}$ is the density matrix of the environment,
which we assume to be in equilibrium at temperature $T$. Under the
assumption of separability of the initial density matrix, we
consider that the system evolution immediately follows an
initialization in a definite state, implemented, e.g., through a
state measurement.

If $\Delta$ is the smallest energy scale in the problem, we can approximate
$U_I(t,0)$ by performing a
perturbation expansion in $\Delta$, which up to second order reads
 \ba
\lefteqn{ U_I(t,0) \approx 1 +\frac{ i}{2} \int_{0}^t dt' \Delta(t')
 \tilde{\sigma}_x(t')}\nonumber\\
 &&-\frac{1}{4}\int_0^t \int_0^{t'}dt'dt''\Delta(t')\Delta(t'')
 \tilde{\sigma}_x(t') \tilde{\sigma}_x(t'').\label{UI}
 \ea
If the time interval $t$ is not small enough to make the above
integrals small, i.e., $t \gtrsim 1/\Delta$, the higher order terms
in $\Delta$ must be considered in the expansion.

\subsection{Off-diagonal elements of $\rho$}

To zeroth order in $\Delta$, we have $U_I(t,0)=1$, therefore
 \ba
 \rho_{SB}(t) =e^{-iH_B t} U_0(t)\rho_{SB}(0)U_0^\dagger(t)e^{iH_B t}.
 \ea
For this case, since $[U_0(t), \sigma_z]=0$, the $\sigma_z$
populations of the system reduced density matrix $\rho$ are
constants of motion. Therefore, in the representation of the
eigenstates of $\sigma_z$,
$\sigma_z\ket{0}=-\ket{0}$($\sigma_z\ket{1}=\ket{1}$), only the
off-diagonal elements of $\rho$ present dynamics, which, due to the
coupling to environment, decay in time. This case constitutes the
one of a pure dephasing dynamics. It has been subject of interest
for many areas, where several approaches have been used to calculate
the off-diagonal elements of $\rho$. Few examples are the (a)
spin-boson model assuming a power-law behavior for the spectral
density of the bath\cite{palma,reina}; (b)  spin-fermion
model\cite{grishin,lutchyn}, and (c) spin-two-state fluctuators
system\cite{schriefl}. Here, we consider a bosonic bath, but do not
have to specify the form of the bath spectral density. To quantify
this decay, let us write the reduced density matrix as
 \be
 \rho(t) = \sum_{i,j=0,1}\rho_{ij}(t)|i\rangle\langle j|.
 \ee
We find for the off-diagonal element
 \ba
\lefteqn{\rho_{01}(t) = \text{Tr}_B[\langle 0| U_0(t)\rho_{SB}(0)U_0^{\dagger}(t)|1\rangle]=\rho_{01}(0)}\nonumber\\
&\times& e^{-i \int_0^t \epsilon(t')dt'}
 \left<\overleftarrow{{\cal T}} e^{-\frac{i}{2}
 \int_0^t Q(t')dt'}{\cal T} e^{-\frac{i}{2}
 \int_0^t Q(t')dt'} \right> ,\label{rhooff}\qquad
 \ea
where $\langle ... \rangle \equiv$ Tr$_B[\rho_B ...]$ and
$\overleftarrow{{\cal T}}$ represents the reverse time ordering
operator. We expand the exponentials, group those in the same order,
take the average of each term, and bring them back to the exponent.
Because of the Gaussian nature of the environment, it is sufficient
to expand up to second order in $Q$. Assuming the environment is in
equilibrium, one finds
 \ba
\lefteqn{  \left<\overleftarrow{\cal T} e^{-\frac{i}{2}
 \int_0^t Q(t')dt'}{\cal T} e^{-\frac{i}{2}
 \int_0^t Q(t')dt'} \right> =}\nonumber\\
 &&= e^{- {1\over 2}\int_0^t dt'\int_0^{t} dt''\left< Q(t')Q(t'') \right> } \nn
 &&= e^{- {1\over 2} \int {d\omega \over 2\pi}
 \int_0^t dt'\int_0^{t} dt''e^{i\omega(t''-t')}S(\omega)}.\label{gaussaverage}
 \ea
Thus, using (\ref{gaussaverage}) in (\ref{rhooff}), we obtain
 \ba
 \lefteqn{\rho_{01}(t)= e^{-i\int_0^t \epsilon(t') dt'}} \nonumber\\
&\times& \exp \left\{ - \int {d\omega \over \pi}
 S(\omega){\sin^2(\omega t/2) \over \omega^2} \right\}\rho_{01}(0).\quad
 \ea
This equation represents a complicated decay rate, which is in
general not a simple exponential function of $t$. However, in two
limits it can be simplified. First, for the case of white noise,
i.e., $S(\omega)=S(0)$, it gives
 \ba
 \rho_{01}(t) = e^{-i\int_0^t \epsilon(t') dt'
 - {1\over 2}S(0)t}\rho_{01}(0).
 \ea
Which leads to dephasing rate $1/T_2 = {1\over 2}S(0)$, as expected
for white noise.

Another interesting limit is when $S(\omega)$ is
dominated by low frequencies so that one can use $\sin x \approx x$
to get
 \ba
 \rho_{01}(t)=  e^{-i \int_0^t \epsilon(t') dt'
 - {1\over 2}W^2 t^2}\rho_{01}(0),
 \ea
where $W$ is the energy level broadening given by (\ref{W}). The
decay is now a Gaussian, whose width determines the dephasing rate,
$1/T_\phi=W$. For the case of 1/f noise, where the cutoff of
$S(\omega)$ is not sharp enough, one gets a logarithmic correction
to the above equation \cite{schriefl}.


\subsection{Diagonal elements of $\rho$}

The evolution of the diagonal part of the density matrix happens in
a time scale much larger than $1/W$. The complete evolution is given
by
 \ba
 \rho(t) = \text{Tr}_B[U_0(t)
 U_I(t,0) \rho(0) \rho_B U_I^\dagger(t,0)U_0^\dagger(t)].
 \ea
Let us assume the initial condition $\rho(0) = |0\rangle\langle 0|$
and try to calculate $\rho_{11}(t)$. To zeroth order in $\Delta$, we
have $\rho_{11}(t)=0$ as expected, thus we find that the first
nonzero contribution to $\rho_{11}(t)$ comes from the first-order
term in $\Delta$ of (\ref{UI}):
 \ba
 \lefteqn{\rho_{11}(t) \approx
 {1 \over 4}\int_0^t d t_1\int_0^t dt_2 \Delta(t_1)\Delta(t_2)}
 \nn &\times&
 \text{Tr}_B[\langle 1|\tilde{\sigma}_{x}(t_1)|0\rangle \rho_B
 \langle 0|\tilde{\sigma}_{x}(t_2)|1\rangle] \nn
&=& {1 \over 4}\int_0^t d t_1\int_0^t
 dt_2 \Delta(t_1)\Delta(t_2) \nn &\times&
 \text{Tr}_B[\langle 1|U_0^\dagger(t_1)U_0^*(t_1)|1\rangle
 \rho_B \langle 0|U_0^{\dagger}(t_2)U_0^*(t_2)|0\rangle] \nn
 &=& {1 \over 4}\int_0^t d t_1\int_0^t
 dt_2 \Delta(t_1)\Delta(t_2) e^{i\int_{t_1}^{t_2} \epsilon(t')dt'} \nn
 &\times&\left< \overleftarrow{{\cal T}} e^{{i\over 2} \int_0^{t_2} Q(t')dt'}
 {\cal T} e^{{i\over 2} \int_0^{t_2} Q(t')dt' } \right. \nn
 && \left. \overleftarrow{{\cal T}} e^{ -{i\over 2}\int_0^{t_1} Q(t')dt'}
 {\cal T} e^{-{i\over 2} \int_0^{t_1} Q(t')dt' } \right>,\label{rhodiagonal}
 \ea
where in the second equality we have used $\tilde{\sigma}_x(t) =
U_0^\dagger(t)\sigma_xU_0(t) = U_0^\dagger(t)U_0^*(t)\sigma_x$. One
can calculate the expectation value by expanding the exponentials
and keeping only the the second order terms. The last two lines of
(\ref{rhodiagonal}) become
 \ba
 \lefteqn{1 + {1\over 2} \int_{0}^{t_2} dt'\int_{t_2}^{t_1} dt''\left< Q(t')Q(t'')
 \right> }\nn
 &&+ {1\over 2} \int_{t_1}^{t_2} dt'\int_{0}^{t_1} dt''\left< Q(t')Q(t'') \right>.
 \ea
Substituting the inverse Fourier transformation $\left< Q(t')Q(t'')
\right> {=} \int {d\omega \over 2\pi} e^{-i\omega(t' - t'')}
S(\omega)$, we find
 \ba
 &&1+ \int {d\omega \over 2\pi}
 {S(\omega)\over \omega^2} [e^{i\omega(t_1-t_2)}-1+i(\sin \omega
 t_2 -\sin \omega t_1)] \nn
 &&= 1+ \int {d\omega \over 2\pi}
 {S(\omega)\over \omega^2} (\cos \omega \tau -1) \nn
 &&- i\int {d\omega \over 2\pi}{S(\omega)\over \omega^2}(\sin \omega \tau
 - 2 \sin {\omega \tau\over 2} \cos \omega \tau'),
 \label{preepsilon}
 \ea
where $\tau = t_2-t_1$ and $\tau' = (t_1+t_2)/2$.

If the noise spectral density $S(\omega)$ is dominated by low
frequency noise such that for all relevant modes $\omega \tau \ll
1$, one can expand the $\sin \omega \tau$ and $\cos \omega \tau$ in
(\ref{preepsilon}) to get
 \ba
  \rho_{11}(t) &\approx&  {1 \over 4}\int_{0}^{t}d\tau' \int_{-\tilde{t}}^{\tilde{t}}d\tau
 \Delta(\tau'+{\tau\over 2})\Delta(\tau'-{\tau\over 2}) \nn
 &\times& e^{- W^2\tau^2/2- i\left(\epsilon_p(\tau')\tau -\int_{-\tau/2}^{\tau/2}\epsilon (\tau'+t')dt'\right)},\label{rhodouble}
 \ea
where $\tilde{t} = \min [2\tau',2(t-\tau')]$, $W$ is given by (\ref{W}), and
 \ba
 \epsilon_p(t) \equiv \int {d\omega \over 2\pi}
 {S(\omega)\over \omega} (1{-} \cos \omega t). \label{epti}
 \ea
Equation (\ref{rhodouble}) conveys the non-locality in time,
expected for a non-Markovian environment. If within time $\tau \sim
1/W$, $\epsilon(t)$ and $\Delta(t)$ do not change much (or even if
$\Delta(t)$ is a fast but linear exponential function), we can write
(\ref{rhodouble}) as
 \be
 \rho_{11}(t) \approx {1 \over 4}\int_{0}^{t}d\tau' \Delta^2(\tau')
 \int_{-\tilde{t}}^{\tilde{t}}d\tau\ e^{i[\epsilon(\tau')-\epsilon_p(\tau')] \tau
 -W^2\tau^2/2}. \label{rho11-0}
 \ee
Therefore, for $t\lesssim1/\Delta(t)$, we find the leading term for
system population rate change given by
 \ba
 \dot \rho_{11}(t) \approx {\Delta^2(t) \over 4}
 \int_{-t}^{t}d\tau \ e^{i[\epsilon(t)-\epsilon_p(t)] \tau
 -W^2\tau^2/2}. \label{rho11-1}
 \ea
If $t > 1/W$, due to gaussian envelope of the integrand, we can
extend the integration limits of (\ref{rho11-1}) to $\pm\infty$,
obtaining
 \ba
 \dot \rho_{11}(t) \approx \Gamma_p \
 e^{-[\epsilon(t) - \epsilon_p(t)]^2/2W^2}, \label{rho11-2}
 \ea
with the peak value of the functions given by
 \be
 \Gamma_p \equiv \sqrt{\pi \over 8}{\Delta^2 \over W}. \label{peak}
 \ee
It is worth recalling that for times $t\gtrsim 1/\Delta$,
Eq.~(\ref{UI}) does not represent a fair approximation to
$U_I(t,0)$, hence the corrections to
Eqs.~(\ref{rho11-0}-\ref{rho11-2}), due to higher powers of
$\Delta$, become appreciable and must be considered.

In section \ref{eqmotion}, we present a detailed study for the
general equation of motion of the reduce density matrix consistent
with (\ref{rho11-2}). However, before we reach that stage, it is
worth discussing a simpler system with a time independent
Hamiltonian, and deriving some general features for $\epsilon_p(t)$
behavior.

\section{Macroscopic resonant tunneling}
\label{equilibrium}

Should $\epsilon_p$ be constant in time and the Hamiltonian
(\ref{HS}) be time independent, one could directly read
(\ref{rho11-2}) as an approximation for the equation of motion
 \be
 \dot{\rho}_{11}(t)\approx \Gamma_-\rho_{00}(t)-\Gamma_+\rho_{11}(t),
 \ee
since the off-diagonal elements of $\rho(t)$ become negligible for
times $t\gtrsim1/W$. The rate $\Gamma_-$ ($\Gamma_+$) then
represents the $\ket{0}\rightarrow\ket{1}$
($\ket{1}\rightarrow\ket{0}$) system transition rate. Thus, for the
regime $1/W\lesssim t\lesssim1/\Delta(t)$, the evolution is
described by (\ref{rho11-2}). The same argument holds when
$\epsilon_p(t)$ is a function of time, but in that case the
tunneling rates will be time dependent:
 \ba
 \Gamma_\pm (t) = \Gamma_p \
 e^{-[\epsilon \pm \epsilon_p(t)]^2/2W^2}. \label{Gammapmt}
 \ea

An experimental realization of such a tunneling process in a
macroscopic quantum device such as a superconducting flux qubit is
called macroscopic resonant tunneling (MRT). The tunneling rates
$\Gamma_\pm$ are therefore simple shifted Gaussian functions
described by (\ref{Gammapmt}). An immediate consequence of
(\ref{epti}) is that the shift $\epsilon_p$ vanishes for a classical
noise, for which $S(\omega)$ is symmetric. Therefore, a nonzero
value of $\epsilon_p$ is a signature for quantum nature of the noise
source.

If the environmental source is in equilibrium at temperature $T$,
then the symmetric and antisymmetric (in frequency) parts of the
noise intensity are related by the fluctuation-dissipation theorem:
 \be
 S_s(\omega)= S_a(\omega) \coth \left({\omega \over 2T}\right)
 \label{FDTheorem}
 \ee
Therefore the fluctuation-dissipation theorem relates $W$ and
$\epsilon_p(t)$, which are functions of $S_s$ and $S_a$ respectively.
Let us first define
 \ba
 \epsilon_{p0} = {\cal P}\int {d\omega \over 2\pi}
 {S(\omega)\over \omega}, \label{ep0}
 \ea
with ${\cal P}$ representing principal value integral. In the case
of low-frequency noise, when all the relevant frequencies are small
on the scale of temperature $T$, i.e., $\omega \ll T$,  one can
write $\coth (\omega/2T)\simeq 2T/\omega$. In that case (\ref{W}),
(\ref{FDTheorem}), and (\ref{ep0}) yield
\begin{equation} W^2 = 2T  \epsilon_{p0} \, . \label{e17} \end{equation}
One therefore finds
 \ba
 \epsilon_p(t) = \epsilon_{p0} - {\cal P}\int {d\omega \over 2\pi}
 {S(\omega)\over \omega} \cos \omega t. \label{ept}
 \ea
The effect of the last term depends on how small or large $t$ is
with respect to the time response, $\tau_R\sim\omega_c^{-1}$, of the
environment. Here, $\omega_c$ represents the characteristic energy
of the environment. To understand this let us consider different
regimes.

\subsection{Large $\omega_c$ (short $\tau_R$) limit}

If $\omega_c$ is large compared to $1/t$, where $t$ is the typical
time scale of interest, then the integral in (\ref{ept}) covers many
oscillations of the cosine function and therefore vanishes. In that
case
 \be
 \epsilon_p \approx\epsilon_{p0} = {W^2\over 2T},
 \ee
consequently being {\it independent} of $t$. For a time independent
Hamiltonian, Eq. (\ref{Gammapmt}) then yields
 \ba
 \Gamma_\pm (\epsilon) = \Gamma_p\ e^{-[\epsilon \pm \epsilon_{p0}]^2/2W^2}, \label{G0110}
 \ea
in agreement with Ref.~\onlinecite{aa08}. It is easy to see that
 \be
 {\Gamma_{-}(\epsilon) \over \Gamma_{+}(\epsilon)} = e^{\epsilon/T},
 \ee
which (in the limit $\Delta {\to}\, 0$) is the detailed balance
(Einstein) relation. Therefore, the transition rates (\ref{G0110})
support thermal equilibrium distribution of the system states, which
is a natural consequence of the fast environmental response.

\subsection{Small $\omega_c$ (long $\tau_R$) limit}

If the environment's response is slow compared to time scale of the
problem, i.e., $\omega_c\ll 1/t$, the cosine function in (\ref{ept})
will be close to 1 at all times, making $\epsilon_p \approx 0$,
again {\it independent} of $t$. For a time independent Hamiltonian,
therefore, we get
 \ba
 \Gamma_{-} = \Gamma_{+} = \Gamma_p\
 e^{-\epsilon^2/2W^2}. \label{G0110-0}
 \ea
Such transitions obviously do not satisfy the detailed balance
relation and do not lead to equilibrium distribution.

Indeed, an environment in $\omega_c \to 0$ regime behaves as a
static (classical) noise source. To see this, let us consider
Hamiltonian (\ref{HS}) with a static noise source $Q$ that does not
vary much during the evolution and has a Gaussian distribution:
 \be
 P(Q) = {e^{-Q^2/2W^2} \over \sqrt{2\pi}W}.
 \ee
In small $\Delta$ regime, the tunneling rate from state $|i\rangle$
to state $|j\rangle$ can be calculated using the Fermi Golden rule
 \be
 \Gamma_{i \to j} = 2 \pi |\langle i|V|j \rangle|^2
 \delta(E_i-E_j),
 \ee
where $V=\Delta \sigma_x/2$ is the perturbation potential. Therefore,
for every realization of $Q$, one finds
 \be
 \Gamma_{-}(Q) = \Gamma_{+}(Q) = {\pi \Delta^2 \over 2} \delta (\epsilon
 + Q)
 \ee
Averaging over all possibilities of Q, we find
 \ba
 \Gamma_{-} = \Gamma_{+} &=& {\pi \Delta^2 \over 2} \int dQ P(Q) \delta
 (\epsilon + Q) \nn
 &=& \Gamma_p\ e^{-\epsilon^2/2W^2},
 \ea
which is the same as (\ref{G0110-0}).

\subsection{General $\omega_c$ $(\tau_R)$ regime}

In general, away from the above two limits, $\epsilon_p(t)$ is a time
dependent function given by (\ref{ept}). The explicit functionality
depends on the spectral density $S(\omega)$, especially on its
characteristic frequency $\omega_c$. To see this, let us assume a simple
spectral density
 \be
 S(\omega) = {2\eta \omega \over [1+(\omega/\omega_c)^2]^2}
 \left(\frac{1}{1-e^{-\omega/T}}\right), \label{SA}
 \ee
for which analytical solutions is possible. Substituting (\ref{SA})
in (\ref{ept}), we find
 \be
 \epsilon_p(t) = \int {d\omega \over 2\pi}
 {\eta (1-\cos \omega t) \over [1+(\omega/\omega_c)^2]^2}
 ={\eta \omega_c \over 4} [1-e^{-\omega_ct}(1+t\omega_c)]
 \ee
We can therefore write
 \ba
 \epsilon_p(t)= \epsilon_{p0}(1-e^{-\omega_ct}(1+t\omega_c)) = \left\{
 \begin{array}{cc} 0,  & \ \ \!\! \omega_ct\ll 1 \\
 \epsilon_{p0},  & \ \ \!\! \omega_ct \gg 1 \end{array} \right.,\quad
 \ea
which yields the above two limiting results in the appropriate
regimes with an exponential crossover between the two limits.
Indeed, the above behavior of $\epsilon_p(t)$, i.e., the crossover
from 0 to $\epsilon_{p0}$ within time scale $\sim 1/\omega_c$, is
generic regardless of the functional detail of $\epsilon_p(t)$. The
time scale $\tau_R \sim 1/\omega_c$ represents the response time of
the environment to an external perturbation. If $t\gg \tau_R$, then
the environment has enough time to enforce equilibrium to the
system, resulting in $\epsilon_p = \epsilon_{p0}$, which is required
for detailed balance (i.e., equilibrium) condition. On the other
hand, if $t\ll \tau_R$, the environment cannot respond quickly to
the system and the equilibrium relation is not expected. In that
case, we find $\epsilon_p = 0$, i.e., the environment behaves as a
classical noise. In the next section we shall see how such behavior
results in time-nonlocality of the equation of motion.

\section{Non-Markovian equation of motion}
\label{eqmotion}

Equation (\ref{rho11-2}) gives the short time ($1/W\lesssim t
\lesssim1/\Delta$) evolution of the diagonal part of the density
matrix. As soon as $t$ becomes comparable to $\Delta$, higher order
corrections become important and the second order perturbation used
in Eq. (\ref{rho11-2}) becomes insufficient. Instead of introducing
higher order corrections which is a cumbersome task, in this section
we take a different path: We write a general equation of motion
expected for the evolution of the density matrix for a system like
ours and find its parameters in such a way that it agrees with
Eq.~(\ref{rho11-2}) for short times.

%
In general, the equation of motion for the evolution of the density
matrix is nonlocal in time, reflecting the non-Markovian nature of
the environment. Since the off-diagonal elements vanish very quickly
(within $t \sim 1/W$), for time scales larger than $1/W$, one can
write the dynamical equations only in terms of the diagonal elements
of $\rho$. Generally, for a non-Markovian dynamics the equation of
motion for $\rho(t)$ depends on the history
 \ba
 \dot \rho_{11}(t) {=} \int_{-\infty}^t dt'[K_{-}(t,t')\rho_{00}(t')
 {-} K_{+}(t,t')\rho_{11}(t')], \label{nlocal}
 \ea
where $K_\pm(t,t')$ are nonlocal integration kernels. Let us from
now onwards consider a time-invariant Hamiltonian for which
 \ba
 \dot \rho_{11}(t) {=} \int_{-\infty}^t dt'[K_{-}(t{-}t')\rho_{00}(t')
 {-} K_{+}(t{-}t')\rho_{11}(t')]. \label{nlocal0}
 \ea
If the system starts the evolution from state $|0\rangle$ at time
$t=t_0$, the short time evolution is described by
 \ba
 \dot \rho_{11}(t) \approx\int_{t_0}^t dt' K_{-}(t{-}t') = \int_{0}^{t-t_0} d\tau
 K_{-}(\tau).
 \ea
This should agree with (\ref{rho11-2}), therefore
 \ba
 \int_{0}^{t-t_0} d\tau K_\pm(\tau) = \Lambda_\pm(t-t_0) \theta(t-t_0).\label{Kintegral}
 \ea
where we have defined functions
 \ba
 \Lambda_{\pm}(t) \equiv \Gamma_p \
 e^{-[\epsilon \pm \epsilon_p(t)]^2/2W^2}. \label{Gammatilde}
 \ea
The presence of the $\theta$-function is necessary to ensure
causality to the system dynamics, since we assume that the evolution
follows a state initialization at $t_0$. Taking the derivative of
both sides of (\ref{Kintegral}), we find
 \be
 K_{\pm}(\tau) = {\partial \Lambda_{\pm}(\tau)\over \partial \tau}
 \theta(\tau) + \Lambda_{\pm}(\tau) \delta(\tau). \label{Kij}
 \ee
Notice that for constant transition rates $\Lambda_{\pm}(\tau) =
\Gamma_{\pm}$, Eq. (\ref{Kij}) leads to
 \ba
 \dot \rho_{11}(t) = \Gamma_{-} \rho_{00}(t) - \Gamma_{+}
 \rho_{11}(t), \label{rholocal}
 \ea
which, as expected, is $t$-local.

In the limit $\omega_c \to 0$, where the change in $\Lambda_{\pm}$
happens on a time scale ($1/\omega_c$) much larger than the time
evolution considered here, the time derivative in (\ref{Kij}) can be
neglected and one obtains (\ref{rholocal}) with transition rates
$\Gamma_{\pm} = \Lambda_{\pm}(0)= \Gamma_p\, e^{-\epsilon^2/2W^2}$,
with $\epsilon_p(t)=0$, as expected for a static noise.

On the other hand, in the $\omega_c \to \infty$ limit, variations of
$\Lambda_{\pm}(t)$ happen in a very short time, hence $\partial
\Lambda_\pm (\tau)/\partial\tau\rightarrow 0$ for $t \gtrsim \tau_R
\sim 1/\omega_c$. Therefore the $t'$-integration in (\ref{nlocal0})
is basically between $t{-}\tau_R$ and $t$. If within this short
range $\rho(t')$ does not change much, one can bring it outside the
integral. In that case, (\ref{nlocal0}) leads to (\ref{rholocal})
with $ \Gamma_{\pm} = \Lambda_{\pm}(t\rightarrow\infty) = \Gamma_p\,
e^{-(\epsilon\pm\epsilon_{p0})^2/2W^2}$, with
$\epsilon_p(t)=\epsilon_{p0}$, which is expected in the detailed
balance regime.

Both of the above regimes led to $t$-local equations for the
diagonal part of the density matrix. However, for finite $\omega_c$,
in general, one gets a nonlocal equation in time. If the system
evolution is slow compared to the time scale $\tau_R \sim
1/\omega_c$, one can substitute the Taylor expansion $\rho_{ij}(t')=
\rho_{ij}(t) + (t'-t)\dot \rho_{ij}(t)$ into (\ref{nlocal0})
obtaining
 \ba
 \dot \rho_{11}(t) = \Lambda_{-}(t) \rho_{00}(t) - \Lambda_{+}(t) \rho_{11}(t) \nn
 +\dot \rho_{11}(t) \int_{t_0}^t dt'{\partial \Lambda(t{-}t')
 \over \partial t} (t{-}t') ,
 \ea
where $\Lambda(t)= \Lambda_{-}(t)+\Lambda_{+}(t)$. Solving for $\dot
\rho_{11}(t)$, one finds (\ref{rholocal}) with transition rates
 \be
 \Gamma_{\pm} = {\Lambda_{\pm}(\infty) \over
 1 {-} \int_0^{\infty} d\tau \tau {\partial \Lambda (\tau)
 / \partial \tau}}= {\Lambda_{\pm}(\infty) \over
 1 {-} \int_0^{\infty} d\tau [\Lambda(\infty) {-}
 \Lambda(\tau)]},\label{integrand}
 \ee
where in the last step we have used integration by parts. The
integral limit was taken to infinity, since the integrand very
quickly vanishes for $\tau \gtrsim1/\omega_c$. All the nonlocal
behavior is captured in the denominator of (\ref{integrand}).
The integrand (\ref{integrand}) is maximum at $\tau=0$, but very quickly vanishes
within $\tau \sim 1/\omega_c$, hence
 $
 \int_0^{\infty} d\tau [\Lambda(\infty) - \Lambda
 (\tau)] \sim [\Lambda(\infty) - \Lambda(0)]/\omega_c\, ,
 $ 
leading to
 \be
 \Gamma_{\pm} \approx {\Lambda_{\pm}(\infty) \over
 1 {-} [\Lambda(\infty) - \Lambda(0)]/\omega_c}.\label{nlGamma}
 \ee
Using (\ref{Gammatilde}), we obtain
 \ba
 && \Lambda(\infty) - \Lambda(0)
 = \Gamma_p
 (e^{-(\epsilon-\epsilon_{p0})^2/2W^2} \nn
 && + e^{-(\epsilon+\epsilon_{p0})^2/2W^2}-2e^{-\epsilon^2/2W^2})\nn
 && = 2\Gamma_p \ e^{-\epsilon^2/2W^2}
  \left(e^{-\epsilon_{p0}^2/2W^2}\cosh {\epsilon \over 2T} -1 \right).
 \ea
Therefore, to the lowest order in $\Gamma_p/\omega_c$, we get
 \ba
 \Gamma_{\pm}(\epsilon) &\approx& \Gamma_p \ e^{-(\epsilon \pm \epsilon_{p0})^2/2W^2}
 \left\{ 1 + {2 \Gamma_p \over \omega_c}\ e^{-\epsilon^2/2W^2} \right. \nn
  && \left. \left(e^{-\epsilon_{p0}^2/2W^2}\cosh {\epsilon \over 2T} -1 \right)
  \vphantom{1\over 2} \right\}. \label{GammapmNL}
 \ea
The magnitude and the position of the peak of $\Gamma_{-}(\epsilon)$
are given by (to the lowest order in $\Gamma_p/\omega_c$)
 \ba
 \Gamma_{\rm peak} &\approx& \Gamma_{-}(\epsilon_{p0}) \approx \Gamma_p
 (1+\Gamma_p/\omega_c), \nn
 \epsilon_{\rm peak} &\approx& \epsilon_{p0} \left(1+{2\Gamma_p \over
 \omega_c}\ e^{-\epsilon_{p0}^2/2W^2} \right).
 \ea
The peak value is enhanced by the nonlocal effects. The peak
position is also shifted, but by a very small amount due to the
exponential suppression. Notice that the peak becomes asymmetric
around its center due to the nonlocality.

The nonlocal corrections to the transition rates become negligible
when $\Gamma_p \ll \omega_c$. Also, observe that $\Gamma_p$ is
approximately the peak value of the transition rate (\ref{peak}).
Therefore, nonlocality becomes important only when the maximum
transition rate $\Gamma_p$ is of the order of or larger than
$\omega_c$, or equivalently, when the response time ($\tau_R$) of
the environment is comparable or longer than the system transition
time ($\sim 1/\Gamma_p$).

\section{MRT in a double-well potential}
\label{doublewell}

So far we have studied incoherent tunneling in an idealized two
state system. However, for most realistic systems, the two state
model is only an approximation of a more complicated multi-level
problem. An example of such cases is a system in which the classical
potential energy has a double-well structure and the kinetic part of
the Hamiltonian provides quantum tunneling between the two wells.
Experimental implementation of such a system is possible using
superconducting flux qubits, which have been studied considerably
both theoretically and
experimentally\cite{super,bkd,lukens,chiorescu,clarke,koch,planteberg,Han95,harrisprl,douglas,janprb}.
Especially, MRT measurements have been performed both between ground
states as well as excited states of the wells.
\cite{Han95,harrisprl}

In such a double-well system, the energy within each well is
quantized, with energy level distributions dependent on the bias
energy between the wells.  In general, in the presence of the
environment, a system initialized in one of those levels can
experience two types of evolution: intra- and interwell dynamics.

The intrawell dynamics are transitions within a single well, e.g.,
when a system excited within a well relaxes to a lower energy level
in the same well by exchanging energy with the environment. Thus, in
this case, the system dynamics is confined in just one well of the
potential, with no tunneling to the opposite well.

It is also possible for the system, depending on the tunneling
amplitude between the two states, to tunnel to an energy level in
the opposite well, leading thus to an interwell dynamics. If the
evolution of the system is confined to the ground states of the two
wells and it lies in the incoherent tunneling regime, then the
formalism developed herein can describe such an evolution in full
detail. This, however, is not the only type of evolution possible
for a double-well system. Here, we also consider possibilities that
the evolution involves the excited states.

\subsection{Tunneling between ground states}

At low enough temperatures, the system can only occupy the lowest
energy states within the wells. In such a case, tunneling can occur
between those energy levels if the levels become in resonance. The
probability of the system being found in state $|1\rangle$ at time
$t$ is given by $\rho_{11}(t)$. For a time independent system
initialized in state $|0\rangle$, in the limit $\Gamma_p \ll
\omega_c$, $\rho_{11}(t)$ is the solution of (\ref{rholocal}). Such
a $t$-dependence can be measured experimentally and is usually an
exponential function with initial value 0 and final value given by
the equilibrium distribution.  According to (\ref{rholocal}), the
initial slope of $\rho_{11}(t)$ gives the transition rate: $\Gamma_-
= \dot \rho_{11}(0)$. Likewise, if the system is initialized in
state $|1\rangle$, one can extract $\Gamma_+$ in a similar way.
Plotting the resulting transition rates versus bias $\epsilon$, one
obtains the tunneling resonant peaks. By fitting the experimental
data to the shifted Gaussian line-shapes (\ref{G0110}) the
parameters $\epsilon_{p0}$, $W$, and $\Gamma_p$ can be extracted and
from (\ref{peak}), $\Delta$ can be obtained. Such a procedure,
performed in Ref.~\onlinecite{harrisprl}, successfully confirmed our
theory especially the relation (\ref{e17}) between $W$ and
$\epsilon_{p0}$.

If the transition rate $\Gamma_p$ becomes comparable to the
environment's characteristic energy $\omega_c$, the $t$-local
equation (\ref{rholocal}) will not be adequate to describe the
evolution of the system. However, if the nonlocality effect is
small, one can still use the same equation but with $\Gamma_\pm$
defined by (\ref{nlGamma}), hence (\ref{GammapmNL}). In such a case,
the peak will not be symmetric around its center, with an asymmetry
that dependens on $\Delta$. Experimental observation of such an
asymmetry is an indication of time-delayed response of the
environment and may provide information about $\omega_c$. It should
be reminded that a presence of high frequency modes in $S(\omega)$
may also lead to deviation from a symmetric Gaussian MRT peak but
such an effect is independent of $\Delta$ hence could be easily
distinguished from the above nonlocal effects.

Another interesting type of experiment is the Landau-Zener
transition in which $\epsilon$ is a linear function of time during
the evolution. For that type of evolution, again in the $\Gamma_p
\ll \omega_c$ regime, one can still use (\ref{rholocal}) but with a
time dependent $\epsilon$. Such a procedure was proved successful in
providing accurate description of the experimental data for flux
qubits in Ref.~\onlinecite{janprb}.

It should be mentioned that the tunneling rate $\Delta$ in our
formalism may not be independent of $\epsilon$ as assumed here. In
practice, as the double-well potential is tilted, it not only
affects the relative positions of the energy levels in the two wells
but also affects the matrix elements between them. Such dependence
is weak for a small bias, but as $\epsilon$ becomes large the effect
of modulation of $\Delta$ might become visible.

\subsection{Tunneling to or between excited states}

If the energy tilt is large enough so that the ground state of the
initial well becomes in resonance with an excited state of the
opposite well, tunneling to the excited state can occur.
Alternatively, one may initialize the system in an excited state in
the initial well, via e.g., microwave excitation, and make the
system tunnel between two excited states. It is therefore important
to understand how such a tunneling can be described within the
present theory. One can generalize the arguments of the previous
section to calculate the tunneling rate. In this case, we need to
add intrawell relaxations to the picture.

In Ref.~\onlinecite{aa08}, it was shown that the tunneling rate from
state $|i\rangle$ in the left well to state $|j\rangle$ in the right
well is given by
 \ba
 \Gamma_{ij}(\epsilon) = {\Delta_{ij}^2 \over 4} \int_{-\infty}^{\infty}dt \
 e^{i(\epsilon - \epsilon_p) t -\gamma_{ij} |t| - {1 \over
 2}W^2t^2},
 \label{GammaExcited}
 \ea
where $\epsilon$ is the bias energy with respect to the resonance
point between $|i\rangle$ and $|j\rangle$, $\Delta_{ij}$ is the
tunneling amplitude between the two states, and $\gamma_{ij} =
(\gamma_i+\gamma_j)/2$, with $\gamma_i$ being the intrawell
relaxation rates corresponding to state $|i\rangle$. If one of the
states is the ground state in its own well, then its corresponding
intrawell relaxation rate is zero. The transition rate becomes a
convolution of Lorentzian and Gaussian functions:
 \ba
 \Gamma_{ij}(\epsilon) = {\Delta_{ij}^2 \gamma_{ij} \over \sqrt{8\pi}W}
 \int_{-\infty}^\infty d\epsilon' {e^{-[\epsilon' - \epsilon_p]^2/2W^2}
 \over (\epsilon - \epsilon')^2 + \gamma_{ij}^2} \nn
 = \sqrt{\pi \over 8}{\Delta_{ij}^2 \over W} \text{ Re}
 \left[w\left({\epsilon \pm \epsilon_p + i\gamma_{ij} \over \sqrt{2}W}
 \right) \right], \label{gengamma}
 \ea
where
 \be
 w(x) = e^{-x^2}[1-\text{erf}(-ix)] = {2e^{-x^2} \over \sqrt{\pi}}
 \int_{ix}^\infty e^{-t^2} dt
 \ee
is the complex error function. In the limit $\gamma_{ij} \to 0$, the
shifted Gaussian line-shape (\ref{G0110}) is recovered. In the
opposite limit, $\gamma_{ij} \gg W$, the peak becomes a Lorentzian
with width $\gamma_{ij}$.

\subsection{Multi-channel tunneling}

So far, we have investigated the dynamics of a definite single
tunneling event between the wells. However, as the system's
temperature increases, one should expect the increase of probability
of thermal occupation of the excited states of each well. Under such
conditions, it becomes unknown what single tunneling event will take
place. Consequently, when predicting the effective tunneling rate
between wells, one has to take into account the statistics of
occupation of excited states and their respective tunneling
probabilities to the opposite well, in an ensemble average. The net
of this thermally assisted dynamics is a multi-channel tunneling,
which leads to an increase of the measured tunneling rate. As we
shall see, due to the fast increase of the tunneling amplitude
$\Delta_{ij}$ between excited states $\ket{i}{\rm and} \ket{j}$, $T$
does not need to be too large for this process to become
non-negligible. For simplicity we consider zero bias ($\epsilon=0$)
situation in which the two potential wells are in resonance.

Let $\Delta_n$ and $\Gamma_\pm^n$ denote the tunneling amplitude and
transition rate between the $n$-th energy levels in the opposite
wells, and $\Gamma_{\pm}$ the {\it total} transition rates between
the wells. In thermal equilibrium, the occupation probability of the
$n$-th state is given by Boltzmann distribution: $P_n =
e^{-E_n/T}/\sum_i e^{-E_i/T}$. Therefore
 \ba
 \Gamma_{\pm}(\epsilon) = \sum_n P_n \Gamma_\pm^n(\epsilon).
 \ea
At small enough $T$, one can assume $P_n \approx e^{-E_{n0}/T}$ (for
$n>0$), where $E_{n0} = E_n - E_0$ is the relative energy of state
$|n\rangle$ compared to the ground state ($n=0$). If $\gamma_{ij}
\ll W$ for the low-lying energy levels, we may neglect $\gamma_{ij}$
in (\ref{gengamma}) and all $\Gamma_n$ will have the same Gaussian
functional form, leading to
 \ba
 \Gamma_{-}(\epsilon) &=& \sum_n e^{-E_{n0}/T} \sqrt{\pi \over 8}{\Delta_n^2 \over W}
 e^{-[\epsilon - \epsilon_p]^2/2W^2}, \nn
 &=& \sqrt{\pi \over 8}{\Delta_{eff}^2(T) \over W}
 e^{-[\epsilon - \epsilon_p]^2/2W^2},
 \ea
where
 \be
 \Delta_{eff} = \Delta_0 \left[ 1 + \sum_{n\geq
 1}{\Delta_n^2\over \Delta_0^2}e^{-E_{n0}/T} \right]^{1/2}.
 \label{Deff}
 \ee
Therefore, the net contribution from tunneling events involving
excited states can be seen as a renormalization of the tunneling
amplitude between wells. Since usually $\Delta_n \gg \Delta_0$, such
contribution becomes important even at temperatures much smaller
than the plasma frequency $\omega_p \equiv E_{10}$. The crossover
temperature $T_{co}$ can be obtained by requiring
$(\Delta_1/\Delta_0)^2e^{-\omega_p/T} \sim 1$, such that the
contribution from the first excited state becomes important:
 \be
 T_{co} = {\omega_p \over 2\ln(\Delta_1/\Delta_0)}.
 \ee
Typically $\Delta_1$ is a few orders of magnitude larger than
$\Delta_0$ and therefore $T_{co}$ can be an order of magnitude
smaller than $\omega_p$. High frequency modes of environment may
also renormalize the tunneling amplitude\cite{leggett} resulting in
a $T$-dependent $\Delta_{eff}$ even at $T<T_{co}$. Such a
$T$-dependence is typically much weaker and a crossover to the
exponential dependence in (\ref{Deff}) should be observable.

\section{Conclusions}
\label{conclusion}

We have shown a systematic procedure to determine the evolution of a
two-state system in the regime of incoherent quantum dynamics.
Considering a second order perturbation theory in the system bare
tunneling rate $\Delta$, and a Gaussian distribution for the
environment fluctuations, we have determined the short time
evolution of the system reduced density matrix elements.

Under the assumption of high integrated noise $W$, i.e., a
system-bath strong coupling regime, we verify that, indeed,
dephasing process takes place early in the system evolution, which
sets $1/W$ as the smallest time scale of the evolution, justifying
the claim of having a system with incoherent dynamics.

As for the system populations, we have seen that, in general, one
should expect complex non-Markovian dynamics. We were able to
clearly demonstrate how the non-Markovian evolution can be related
to the time response of the environment, $\tau_R$. Indeed, we have
verified that for time scales $t\gg\tau_R$, the system follows the
detailed balance dynamics. On the other hand, if the environment
response is very slow, i.e., $t\ll\tau_R$, the system sees a static
(classical) noise source. In addition, by investigating the equation
of motion for the reduced density matrix, we have demonstrated how
one can simplify the non-Markovian effects by introducing modified
transition rates for the dynamical equations.

Finally, we have inspected the intra- and interwell transition
possibilities inside a double-well potential, and quantified how the
multi-channel process can lead to an enhancement of the system
tunneling. We have determined the condition for this process to take
place, and estimated the crossover temperature which can be an order
of magnitude smaller than the system plasma frequency $\omega_p$.

Some of the predictions of our theory have already been confirmed
experimentally.\cite{harrisprl,janprb} More experiments, however,
are necessary especially to confirm our description of non-Markovian
dynamics. A simple measure of the asymmetry of the MRT peak in large
$\Delta$ regime could be indicative of nonlocality in $t$. As
described in Sec.~V, such an asymmetry should be $\Delta$ dependent
and should disappear at small $\Delta$. A $\Delta$-independent
asymmetry could result from high frequency components of the
environmental noise that make small $\omega \tau$ expansion in
(\ref{preepsilon}) fail. Moreover, a $T$-dependent measure of the
tunneling rates can reveal the renormalization of the effective
tunneling amplitude $\Delta$ due to high frequency noise and the
crossover temperature $T_{co}$ to the multichannel tunneling regime
as described in section VI.

\section*{Acknowledgments}

We would like to thank D.V. Averin, A.J. Berkley, R. Harris, J.
Johansson and T. Lanting for useful discussions and a critical
reading of this manuscript.


\begin{thebibliography}{99}
\bibitem{footnote} Recently, it has been argued that the noise produced by a
system of two-state fluctuators can exhibit non-Gaussian
distribution, playing important role in the qubit-fluctuator strong
coupling limit. See, e.g., Ref.~\onlinecite{nongaussian}.
\end{thebibliography}
\end{document}